\documentclass[aps, twocolumn, pra, showpacs, superscriptaddress]{revtex4}

\usepackage{graphicx}
\usepackage{dcolumn}
\usepackage{bm}
\usepackage{amsmath}

\begin{document}

\title{Fermi super-Tonks-Girardeau state for attractive Fermi gases in an optical lattice}
%\title{Properties of super-Tonks-Girardeau state for attractive Fermi gases in an optical lattice}
%\title{Exotic properties of upper-branch state for attractive Fermi gases in an optical lattice}
%\title{Super-Tonks-Girardeau state for attractively interacting fermions in an optical lattice}
\author{Li Wang}
\affiliation{Beijing National Laboratory for condensed matter
physics, Institute of Physics, Chinese Academy of Sciences, Beijing
100190, China}

\author{Zhihao Xu}
\affiliation{Beijing National Laboratory for condensed matter
physics, Institute of Physics, Chinese Academy of Sciences,
Beijing 100190, China}

\author{Shu Chen}
\email{schen@aphy.iphy.ac.cn}
\affiliation{Beijing National
Laboratory for condensed matter physics, Institute of Physics,
Chinese Academy of Sciences, Beijing 100190, China}

\date{}

\begin{abstract}
We demonstrate that a kind of highly excited state of strongly
attractive Hubbard model, named of Fermi super-Tonks-Girardeau
state, can be realized in the spin-1/2 Fermi optical lattice
system by a sudden switch of interaction from the strongly
repulsive regime to the strongly attractive regime. In contrast to
the ground state of the attractive Hubbard model, such a state is
the lowest scattering state with no pairing between attractive
fermions.  With the aid of Bethe-ansatz method, we calculate
energies of both the Fermi Tonks-Girardeau gas and the Fermi
super-Tonks-Girardeau state of spin-1/2 ultracold fermions and
show that both energies approach to the same limit as the strength
of the interaction goes to infinity. By exactly solving the quench
dynamics of the Hubbard model, we demonstrate that the Fermi
super-Tonks-Girardeau state can be transferred from the initial
repulsive ground state very efficiently. This allows the
experimental study of properties of Fermi super-Tonks-Girardeau
gas in optical lattices.

\end{abstract}

 \maketitle
% 03.75Ss degenerate Fermi gases
% 03.75.Kk Dynamic properties of condensates
% 03.75.Lm Tunneling, Josephson effect, Bose¨CEinstein condensates inperiodic potentials, solitons, vortices, and topological excitations
% 37.10.Jk Atoms in optical lattices
% 42.50.-p Quantum optics

\section{Introduction}

Ultracold atomic gases trapped in one-dimensional (1D) waveguides
have become one of the most active research field of cold atom
physics in recent years \cite{rmpbloch}. Due to their good
tunability, the 1D atomic gases have provided an ideal platform
for studying and testing the theory of low-dimensional many-body
systems \cite{paredes04,kinoshita04}.  Tuning the effective
interaction strength between atoms via Feshbach resonance has led
to the experimental realization of Tonks-Girardeau (TG) gases
\cite{paredes04, kinoshita04}, which describes the Bose gas in the
strongly repulsive limit and exhibits the feature of
fermionization. By switching the interaction between atoms of an
initial TG gas from strongly repulsive to strongly attractive, the
experimental observation of a 1D super Tonks-Girardeau (STG) gas
\cite{Astrakharchik1} of bosonic cesium atoms was reported very
recently \cite{Haller}. In contrast with the strongly repulsive
interacting TG gas, bosons in STG state interact via strongly
attractive interaction. A surprising feature of this many-body
state is its good stability even under strongly attractive
interaction instead of decaying into the lower atomic bound states
\cite{mcguire,tempfli,batchelor}. The stability of the STG gas
could be well understood from the quench dynamics of the 1D
integrable Bose gas \cite{schen}.

The experimental realization of stable excited quantum gas phase
opens a new area for searching exotic quantum phases in ultra-cold
systems
\cite{Rosch,Kantian,schen,girardeau10,lw,Girardeau_BBSTG,Girardeau_FSTG,Guan_Chen,Muth,Chen2,Kormos}.
Particularly, the exotic experimental results \cite{Haller} have
stimulated intensive theoretical studies of the STG gases from
various aspects
\cite{schen,girardeau10,lw,Girardeau_BBSTG,Girardeau_FSTG,Guan_Chen,Muth,Chen2,Kormos}.
So far, most of the theoretical works on the STG gases have
focused on the bosonic gases in continuum systems. In this work,
we study the possible realization of the Fermi
super-Tonks-Girardeau (FSTG) gas for a Fermi gas loaded into a
deep 1D optical lattice, which is described by the Fermi Hubbard
model \cite{LiebandWu}. The FSTG state for a 1D Fermi continuum
gas was studied in Ref. \cite{Girardeau_FSTG,Guan_Chen}, however
it is not clear whether the FSTG gas could be also realized in the
lattice systems. In comparison with the continuum Yang-Gaudin
model \cite{Yang,Gaudin}, the Hubbard model exhibits some new
features due to the existence of the band structure for the
lattice model. One of the new features is the existence of
repulsive bound pairs in the high bands which is absent in the
continuum model. The other one is the existence of the Mott
insulating phase in the half-filling case. Although the Hubbard
model is one of the fundamental model in condensed matter physics,
most of previous theoretical works focused on its ground state and
thermodynamical properties. Our study shall shed lights on
properties of some highly excited states which can be accessible
in current experimental conditions.

Stimulated by the experiment of the bosonic STG gas, we suppose
that the ultracold Fermi gas is initially in the strongly
repulsive regime, and then the interaction is suddenly switched to
the strongly attractive regime. Through this way, we can reach a
stable highly excited phase which is the lowest scattering state
of the attractive Fermi gas. To understand properties of the STG
state of the attractive Hubbard model, we shall analyze the
spectrum structure of the Hubbard model for both the repulsive and
attractive cases. By calculating the energy of the FSTG state
analytically based on the Bethe-ansatz (BA) method, we show that
the energy of FSTG gas state in the strongly attractive limit
approaches the same limit of the ground state energy of 1D
strongly repulsive Fermi Hubbard model. This implies that the FSTG
state can be accessible from the the ground state of the strongly
repulsive Fermi gas by a sudden switch of interactions, which is
also confirmed by the exact calculation of the quench dynamics of
the 1D Fermi gas on the optical lattice by using numerical exact
diagonalization method.

\section{Model and FSTG state}

We consider a 1D ultracold Fermi gas composed of
$N=N_{\uparrow}+N_{\downarrow}$ spin-1/2 fermionic atoms in a deep
optical lattice, which can be well described by the well-known
Hubbard model (HM),
\begin{equation}
\hat{H}=
-t\sum_{i,\sigma}\left(\hat{c}_{i,\sigma}^{\dagger}\hat{c}_{i+1,\sigma}+
\hat{c}_{i+1,\sigma}^{\dagger }\hat{c}_{i,\sigma} \right) +U \sum_i
\hat{n}_{i\uparrow}\hat{n}_{i\downarrow} , \label{HM}
\end{equation}
where $\hat{c}_{i,\sigma}^{\dagger }$($\hat{c}_{i,\sigma}$) with
$\sigma=\uparrow, \downarrow$ is the creation (annihilation)
operator of fermions at the $i$th site, $ t $ and $U$ denote the
hopping amplitude and the on-site interaction strength,
respectively. The ratio $U/t$ can be tuned by varying the depth of
the optical lattice potential and by using Feshbach resonance
technique. For convenience, we set $t=1$ as the energy scale.
Without loss of generalization, we assume that $N_{\downarrow}\leq
N_{\uparrow}$.

The one-dimensional Hubbard model (\ref{HM}) with periodic
boundary condition  is exactly solvable by the Bethe-ansztz (BA)
method \cite{LiebandWu} with the BA wavefuction
\begin{align}
\phi (x_1,...,x_N)=&\sum_{Q} \sum_{P} \theta(x_{Q1}\le...\le x_{QN})
\times \nonumber \\
&[Q,P] \exp [{i \sum_{j=1}^{N} k_{Pj} x_{Qj}}],
\end{align}
where $k_{i}$s represent quasimomenta, $P$s and $Q$s represent
permutations of $k_i$s and $x_i$s, respectively. For the
eigenstate with the total spin $S=N/2-M$ ($M=N_{\downarrow}$), the
coefficient $[Q,P]$ can be explicitly expressed as
$[Q,P]=\epsilon(Q_1)\epsilon(Q_2)\theta(y_1,y_2,...,y_M)\Psi(y_1,y_2,...,y_M;P)$,
where $Q_1$ is the ordering of the first $M$ fermions and $Q_2$
the ordering of the rest fermions, $y_1,y_2,...,y_M$ are the
coordinates of down spins in the lattice with length $L$, and
$\Psi(y_1,y_2,...,y_M;P)=\sum_{R}\epsilon(R)\prod_{j<l}(\Lambda_{Rj}-\Lambda_{Rl}-iU/2t)
\prod_{i=1}^{M}[\prod_{s=1}^{y_i-1}(\sin
k_{Ps}-\Lambda_{Ri}+iU/4t)\prod_{t=y_i+1}^{N} (\sin
k_{Pt}-\Lambda_{Ri}-iU/4t)]$. The parameters $k_j$s and
$\Lambda_\alpha$s are determined by the Bethe-ansatz equations
(BAEs) \cite{LiebandWu}:
\begin{align}
&k_j L=2\pi I_j-2\sum_{\beta =1}^{M} \tan^{-1} (\frac{\sin
k_j-\Lambda_{\beta}}{U/4t}), \label{BAEa}\\
\sum_{j=1}^{N} 2 & \tan^{-1}( \frac{\Lambda_{\alpha}-\sin
k_j}{U/4t}) =2\pi J_{\alpha}+2\sum_{\beta=1}^{M}\tan^{-1}
(\frac{\Lambda_{\alpha}-\Lambda_{\beta}}{U/2t}  ), \label{BAEb}
\end{align}
where $L$ is the size of the optical lattice. The eigenvalues are
given by $E=-2t\sum_{j=1}^{N} \cos k_j$. The structure of the
solution of BAEs of Hubbard model is relevant to the filling
factor of $N/L$. In the following, we shall consider the case with
$N/L <1$.

The BAEs (\ref{BAEa}) and (\ref{BAEb}) hold true for both the
repulsive and attractive $U$, however the structure of the
solutions is quite different for $U>0$ or $U<0$. For the repulsive
interaction with $U>0$, both the solutions of $k_j$ and
$\Lambda_{\alpha}$ for the ground state (GS) and low excited
states are real numbers. The ground state solution corresponds to
$I_j=(N+1)/2-j$ and $J_{\alpha}=(M+1)/2-\alpha$. In the strongly
repulsive interaction limit $U/t \rightarrow \infty$, the ground
state energy is identical to that of a polarized N-fermion system
\cite{Shiba}. On the other hand, if the on-site interaction
between fermions with different spin is attractive, i.e. $U<0$,
the ground state is then composed of $N-2M$ real $k_i$s and $2M$
complex ones. In the strongly attractive interaction limit $U/t
\ll 1 $, the complex solutions take the 2-string form \cite{Penc}:
$\sin k_{\alpha}\approx \Lambda_{\alpha}+i|U|/4t$, and $\sin
k_{M+\alpha} \approx \Lambda_{\alpha}-i |U|/4t$. Besides the
complex solutions, the BAEs also have real solutions for $U<0$,
which describe the scattering states
%and are highly excited states
of attractive fermions. The Fermi super-Tonks-Girardeau (FSTG) gas
state corresponds to the lowest real solutions of BAEs
(\ref{BAEa}) and (\ref{BAEb}) with $U<0$. These scattering states
are gas-like excited states of the attractive spin-1/2 ultracold
fermions which are above states including at least one paired
bound state, while the ground state of the system is composed of
$M$ tightly bound fermion pairs.

\begin{figure}[tbp]
\includegraphics[width=9.5cm]{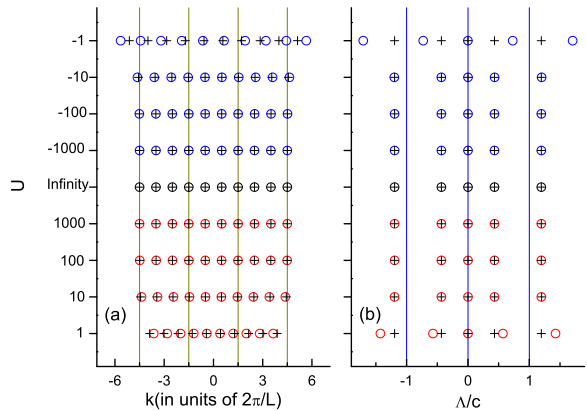}
\caption{(Color online) Comparison between exact solutions
(denoted by $\bigcirc$) and approximate solutions (denoted by $+$)
of BAEs. (a) Quasi-momentum distributions for the ground state of
the repulsive Fermi gas and the FSTG state of the attractive Fermi
gas with different values of $U/t$ and $t$ is set to $1$ as energy
scale here. (b) The corresponding solutions of $\Lambda_{\alpha}$
for different values of $U/t$.} \label{quasik}
\end{figure}

\begin{figure}[tbp]
\includegraphics[width=9.5cm]{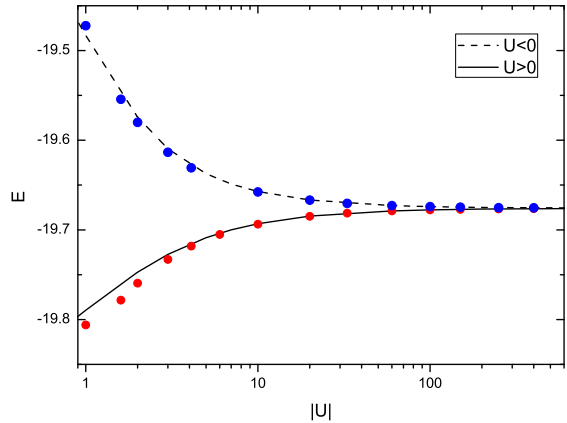}
\caption{(Color online) The energies $E_{FTG}$ (solid line) and
$E_{FSTG}$ (dashed line) vs $U$. Dots in the figure denote the
approximate solutions via expansion in strongly interacting limit.
For large enough $|U|$, they agree very well with the exact
solutions.} \label{energyba}
\end{figure}

Next we explore the scattering solution of the Bethe-ansatz
equations in the strongly interacting limit. As $|U|/t\rightarrow
\infty$, the solution of $\Lambda_\alpha$ is proportional to $U$,
whereas $\sin k_j$ is always finite with $|\sin k_j| \leq 1$.
Therefore the quasimomenta can be given approximately
\begin{equation}
k_j L=2\pi I_j+A_0-A_1 \left( \frac{\sin k_j}{U'}\right)-A_2\left(
\frac{\sin k_j}{U'}\right)^2+O(U'^{-3}) \label{momentum}
\end{equation}
where
\begin{align*}
\left \{
\begin{array}{lrr}
A_0=2\sum_{\alpha=1}^{M} \tan^{-1}\left (\frac{\Lambda_{\alpha}}{U'}
\right)\\
A_1=2\sum_{\alpha=1}^{M}\frac{1}{(\Lambda_{\alpha}/U')^2+1}\\
A_2=2\sum_{\alpha=1}^{M}\frac{\Lambda_{\alpha}/U'}{[(\Lambda_{\alpha}/U')^2+1]^2} \\
\end{array}
\right .
\end{align*}
with $U'=U/4t$. Here we consider the case with
$N_\uparrow=N_{\downarrow}$. Under this condition, the values of
$\Lambda_{\alpha}$ are symmetric about zero. It follows that
$A_0=A_2=0$ and correspondingly Eq.(\ref{momentum}) is simplified
as
\begin{equation}
k_j L=2\pi I_j-A_1 \frac{\sin k_j}{U'}+O(U'^{-3}), \label{kjn2m}
\end{equation}
i.e.,
\begin{align*}
\left \{
\begin{array}{crr}
k_jL=2\pi I_j-\varsigma  \frac{\sin k_j}{|U^{\prime| }} +O\left(
U^{\prime -3}\right)  & \;\; & U>0 \\  k_jL=2\pi I_j+\varsigma
\frac{\sin k_j}{|U^{\prime| }}
+O\left( U^{\prime -3}\right)  & \;\; & U<0 \\
\end{array}
\right . \label{quasikj}
\end{align*}
where $\varsigma=A_1$. In general, $A_1(U)\neq A_1(-U)$ since the
solution $\Lambda_{\alpha}$ of Eq. (\ref{BAEb}) are not symmetric
for $U$ and $-U$. However, in the strongly interacting limit, up
to the order of $U^{-1}$ Eq. (\ref{BAEb}) becomes $2\tan
^{-1}\left( \frac{\Lambda _\alpha }{U^{\prime }}\right) =\frac
1N2\pi J_\alpha +\frac 1N\sum_{\beta =1}^M2\tan ^{-1}\left(
\frac{\Lambda _\alpha -\Lambda _\beta }{2U^{\prime }}\right)$,
which has the same form as BAE of the Heisenberg spin chain and is
invariant under the operation $P:{U\rightarrow-U,
\Lambda_{\alpha}\rightarrow -\Lambda_{\alpha}}$. Therefore, we
have $A_1(U) =  A_1(-U)$ up to the order of $U^{-2}$. It follows
that the ground state energy of the Fermi Tonks-Girardeau (FTG)
gas in the strongly repulsive limit and the energy of FSTG state
in the strongly attractive limit are given by
\begin{eqnarray*}
E_{FTG} &=&-2t\sum_{i=1}^N\cos \left[ \frac{2\pi I_j%
}L\left( 1-\frac \varsigma {L\left| U^{\prime }\right| }\right)
\right]+O(U'^{-3})
 \\
E_{FSTG} &=& -2t\sum_{i=1}^N\cos \left[ \frac{2\pi I_j%
}L\left( 1+\frac \varsigma {L\left| U^{\prime }\right| }\right)
\right]+O(U'^{-3})
\end{eqnarray*}
where $I_j=(N+1)/2-j$ for both the FTG and FSTG gas. Here, for
convenience, we call the ground state of the spin-1/2 Fermi gas in
the strongly repulsive limit as the FTG state. Obviously, in the
limit of $|U|\rightarrow\infty$, we have $E_{FSTG}=E_{FTG}$. In
Fig. \ref{quasik}, we make a comparison between the exact
solutions given by numerically solving the Eq. (\ref{BAEa}) and
(\ref{BAEb}) directly and the approximate solutions given by
solving Eq. (\ref{kjn2m}) iteratively. For an example systems with
$L=100$, $N=10$ and $M=5$, we show that for large enough values of
$U/t$, the approximate solutions agrees very well with the exact
solutions. In Fig. \ref{quasik} (a), one can see that the
quasimomentum distributions for the ground state of repulsive
Hubbard model and the FSTG state approach the same limit from
different sides when $|U|/t$ goes infinite. Correspondingly,
$E_{FSTG}$ and $E_{FTG}$ also approach the same limit as shown in
Fig. \ref{energyba}.

\section{Preparation of FSTG state}
The FSTG state can be realized in a 1D deep optical lattice by a
sudden switch of interaction similar to the experimental
realization of bosonic STG gas in Ref. \cite{Haller}. Suppose that
the initial state $\left|\Psi_{ini}(t=0)\right > =
\left|\psi_0(U_0) \right >$ is prepared at the ground state in the
strongly repulsive regime with $U_0/t \gg 1$, after a sudden
switch to the opposite regime with interaction strength $U/t \ll
-1$, the wave-function $\left| \Psi(t)\right>=e^{-iH(U)t}\left|
\Psi_{ini}(U_0)\right>$ can be calculated via
\begin{equation}
\left| \Psi(t)\right>=\sum_{n} e^{-iE_n t} c_n \left|
\psi_n(U)\right> , \label{psit}
\end{equation}
where $c_n=\left< \psi_n(U)\right| \left. \psi_0(U_0)\right>$ with
$\left|\psi_n (U)\right>$ representing the $n$-th eigenstate of
the Hubbard model with on-site interaction $U$. It is
straightforward that $|c_n|^2$ is the transition probability from
the initial state to the $n$-th eigenstate of $H(U)$.
%To study the quench dynamics of the HM, we shall scrutinize the
%full spectra and eigenstates of the HM numerically by using exact
%diagonalization (ED) method. Although ED can only deal with
%few-body systems due to the limit of computer memory, it gives a
%clear and inspirational picture for understanding the properties
%of the FSTG state.

%\begin{figure}[tbp]
%\includegraphics[width=9.5cm]{fstg03.eps}
%\caption{ Full spectra for the two-particle HM with $L=50,N=2,M=1$,
%$U=10$ (left) and $U=-10$ (right) in $K$ space. The hopping term $t$
%is set to 1.} \label{Ek2}
%\end{figure}

\begin{figure}[tbp]
\includegraphics[width=10cm]{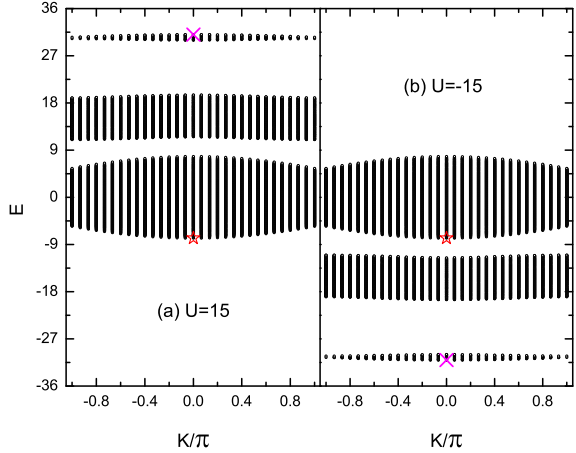}
\caption{ Full spectra of energies vs. total momentum $K$ for the
HM with $L=30,N=4,M=2$, $U=15$ (left) and $U=-15$ (right). The
hopping term $t$ is set to 1. } \label{Ek4}
%Inset on the left bottom is a
%magnified picture of the top band. Inset on the right top is the
%magnified picture of the lowest band.} \label{Ek4}
\end{figure}

\begin{figure}[tbp]
\includegraphics[width=9cm]{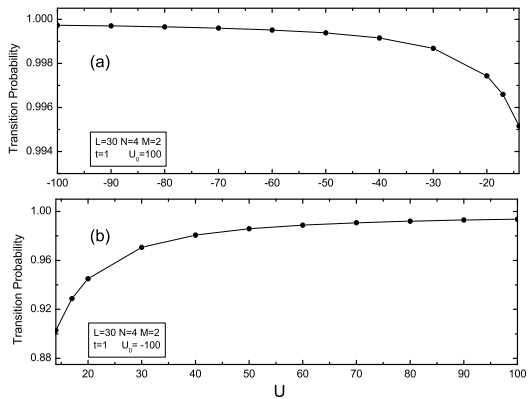}
\caption{ (a) The transition probability from the initial ground
state of the repulsive system with $U_0=100$ to the FSTG state
after the sudden interaction switch to the attractive regime. (b)
The transition probability from the initial ground state of the
attractive system with $U_0=-100$ to the highest excited state of
the top band of repulsive Hubbard model after the sudden switch of
interaction. The results are obtained by numerically exact
diagonalization.} \label{tp}
\end{figure}

\begin{figure}[tbp]
\includegraphics[width=10cm]{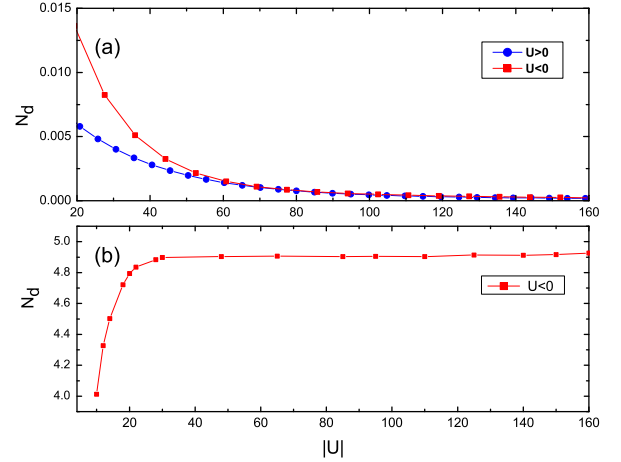}
\caption{ Doubly occupied sites $N_d$ versus the interaction U for
a system with $L=20, N=10, M=5$. (a) STG state for system with
$U<0$ and ground state for system with $U>0$. (b) Ground state for
system with $U<0$. } \label{Eg2}
\end{figure}

To give a concrete example which may help us get an intuitive
understanding of the properties of the FSTG state of the
attractive Hubbard model, we display the full energy-momentum
spectra of the Hubbard model with $L=30$, $N=4$, $M=2$, and $U=\pm
15$ in Fig. \ref{Ek4}. As shown, the spectra is spilt into a
series of separated bands. For the repulsive case, the lowest band
is a scattering continuum of $N$ (here $N=4$) unpaired fermions.
The middle band is a scattering continuum formed by one tightly
bound fermion pair and two unpaired fermions, whereas the top band
is a scattering continuum of two tightly bound fermion pairs. The
spectra for the attractive case is similar but in reverse order.
The gap between centers of neighboring bands equals approximately
to the binding energy $|U|$ of a fermion pair. These separated
bands are no longer distinguishable as the interaction strength is
comparable to the bandwidth. The zero-momentum lowest scattering
state of $N$ unpaired fermions denoted by a red star in Fig.
\ref{Ek4}(a) is just the ground state of the repulsive Fermi gas,
whereas the red star in Fig. \ref{Ek4}(b) indicates the Fermi
super-Tonks-Girardeau gas state.

Starting from the ground state of the HM with a repulsive
interaction and then suddenly switching the interaction to the
attractive side, we evaluate the transition probabilities from the
initial repulsively GS to each eigenstate of the attractive
Hubbard model by the method of the exact diagonalization.  As
shown in Fig. \ref{tp}a, we find that the transition probability
to the lowest state (FSTG) of a given $S$ in the top scattering
band is very close to 1, when both the repulsive interaction of
the initial state and the attractive interaction of the final
state are strong enough. As the transition probability to the
lowest scattering phase is almost $1$ in the strongly interacting
regime, the transition probability to the lower paired states is
almost completely suppressed,  thus we expect that such a highly
excited gas-like state of the strong attractive Fermi gas in
optical lattice can be experimentally realized. Actually, the
stable excited scattering state prepared in this way can be viewed
as a Fermi generalization of the STG gas in the optical lattice.
If the system enters to the weakly interacting regime, the
transition probability to the FSTG state decreases quickly,
whereas the transition probability to the ground state increases.
The FSTG state is more stable when the interaction strength is
closer to the Feshbach resonance point. When the interaction
strength is weak, the FSTG state is not expected to be stable.

Due to the existence of band structure, there exist states of
repulsively bound pairs above the lowest continuum band for the
repulsive Hubbard model. Such kind of repulsively pairing state is
absent in the continuum Yang-Gaudin system \cite{Guan_Chen}.
Particularly, the top band of the strongly attractive Hubbard
model is completely composed of repulsively bound pairs, which is
very similar to the ground state of attractive Hubbard model
composed of attractively bound pairs. Next we show that the
highest excited state composed of repulsive bound pairs can be
realized from the ground state of strongly attractive Hubbard
model by a sudden switch of interaction from $U<0$ to $U>0$.  To
see it clearly, we calculate the transition probability from the
ground state of attractive Hubbard model (marked by the symbol of
cross in Fig. 3b) to the highest state in the top band of the
repulsive Hubbard model (marked by the symbol of cross in Fig.
3a). As shown in Fig. \ref{tp}b, when both the attractive
interaction of the initial state and the repulsive interaction of
the final state are strong enough, the transition probability is
very close to 1. By this way, we can realize the repulsively
paired state for the repulsive Hubbard model. We note that such a
repulsively paired state is a very highly excited state with zero
total momentum, which is different from the $\eta$-pairing state
discussed in Ref.\cite{Kantian,etapairing}.

Next we calculate the number of doubly occupied sites $N_d =
\sum_i \left< \hat{n}_{i\uparrow} \hat{n}_{i\downarrow} \right>$
for the FSTG state by exact diagonalization, which can be obtained
by differentiation of the energy. As shown in Fig. \ref{Eg2}a,
$N_d$ decreases monotonically with increasing $\left|U\right|$ and
tends to zero for $U \rightarrow -\infty$. This indicates that
there is no paring for the FSTG state even in the strongly
attractive limit. As a comparison, we also calculate $N_d$ for the
ground state of the attractive Fermi gas in Fig. \ref{Eg2}b.
Instead, $N_d$ for the ground state of the attractive Fermi gas
monotonically increases to $N/2$ as the ground state is composed
of $N/2$ pairs of fermions with the bounding energy proportional
to $U$. For a bosonic STG gas, it is known that the STG state has
even stronger local correlation than the repulsive TG gas
\cite{Astrakharchik1,schen}. To see whether FSTG state in the
optical lattice has similar properties, we also calculate the
local correlation function for the repulsive ground state. Here we
note that the local correlation function $\sum_i \left<
\hat{n}_{i\uparrow} \hat{n}_{i\downarrow} \right>$ is noting else
but $N_d$ defined above. As shown in in Fig. \ref{Eg2}a, $N_d$ for
the ground state of repulsive Fermi gas also monotonically
decreases to zero with the increase in the repulsion strength $U$.
Nevertheless, the local correlation functions shown in Fig.
\ref{Eg2}a indicate that the FSTG state has stronger local
correlation than the corresponding repulsive ground state, which
is similar to its bosonic correspondence.

\begin{figure}[tbp]
\includegraphics[width=10cm]{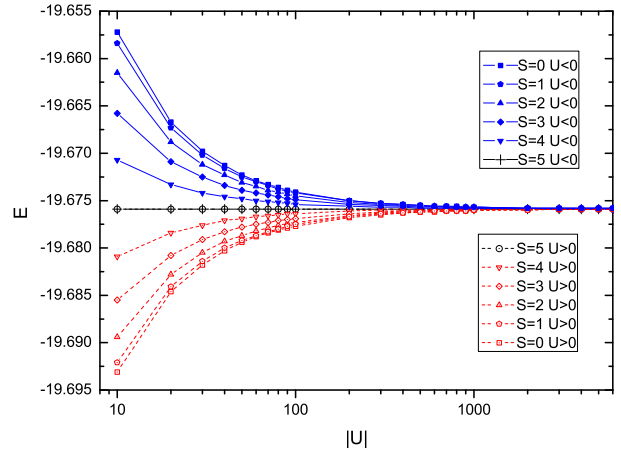}
\caption{ (Color online) Energy vs $U$ for states with different
total spin $S$. } \label{fig-Es}
\end{figure}

In contrast to the spinless Bose system, the ground state of a
spin-$1/2$ Hubbard model is highly degenerate in the limit of $U
\rightarrow \infty$ due to the existence of the spin degree
\cite{FTG}. Nevertheless, the degeneracy is broken for a large but
finite interaction strength and the true ground state of the
repulsive fermions is the state with the lowest $S$
\cite{Lieb-Mattis}. In Fig.{\ref{fig-Es}}, we show the ground
state energy of repulsive Fermi gases and the lowest energy of the
FSTG state for systems with a fixed $N=10$ but various $S$. It is
clear that $E(S_1) < E(S_2)$ for $S_1<S_2$ on the repulsive side.
The energy difference between different spin states vanishes as $U
\rightarrow \infty$. On the other hand, for the FSTG state, the
state with the larger S has lower energy, i.e.,  $E(S_1) > E(S_2)$
for $S_1<S_2$ on the attractive side, as shown in
Fig.{\ref{fig-Es}}. This implies that the ferromagnetic state with
maximum $S=N/2$ has lowest energies for the FSTG states with large
but finite interaction strength. Nevertheless, energies for states
with different total spins approach the same limit of the
polarized state as $|U| \rightarrow \infty$.

\begin{figure}[tbp]
\includegraphics[width=10cm]{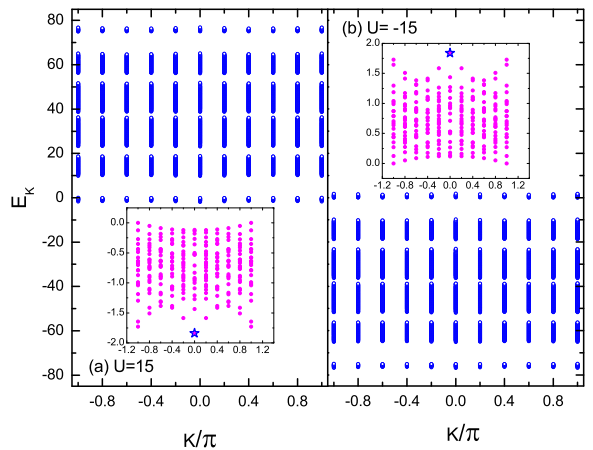}
\caption{ (Color online) Full spectra of energies vs. total
momentum $K$ for the HM with
$L=10,N=10,N_{\uparrow}=N_{\downarrow}=5$, $U=15$ (left) and
$U=-15$ (right). The left inset is the enlargement of the lowest
band of the half-filled repulsive Hubbard model whereas the right
inset is the enlargement of the top band of the half-filled
attractive Hubbard model. } \label{fig7}
\end{figure}

Finally, we discuss the half-filling case, for which the ground
state is a Mott state for any finite repulsion \cite{LiebandWu}.
Similar to the low-density case, we display the full
energy-momentum spectra of the the half-filling Hubbard model with
$L=10$, $N=10$, $N_{\uparrow}=N_{\downarrow}=5$, and $U=\pm 15$ in
Fig. \ref{fig7}. Similarly, the spectra is spilt into a series of
separated bands. For the repulsive case, the lowest band
corresponds to the Mott states with each site occupied by a single
fermion. Above the Mott band, the separated middle bands
correspond to the scattering continuum composed of paired fermions
and unpaired fermions, whereas the top band is the scattering
continuum of $N/2$ paired fermions. In contrast to Fig.3, the
lowest Mott band is obviously very narrow as the hopping process
of a single fermion to its neighboring sites is suppressed. In the
large U limit with $U/t \gg 1$, the double occupied states have
much large energy than the states with no double occupancy, and
the effective Hamiltonian is given by \cite{Duan}
\begin{equation}
H_{eff} = \frac{4t^2}{U} \sum_{i} ( \hat{S}_i \hat{S}_{i+1} -
\frac{1}{4})£¬\label{Heff}
\end{equation}
where $\hat{S}^{\alpha}_i = \frac{1}{2}
\hat{c}_{i,\sigma}^{\dagger}
\sigma^{\alpha}_{\sigma,\sigma'}\hat{c}_{i,\sigma'}$
($\alpha=x,y,z$) are the usual spin operators with
$\sigma^{\alpha}$ being the Pauli matrices. As shown in the inset
of Fig.7a, the enlarged spectrum of the lowest band is consistent
with the spectrum of antiferromagnetic (AFM) Heisenberg model of
Eq.(\ref{Heff}). For the attractive case with $U=-15$,  as shown
in Fig.7b, the spectrum has similar structure as the repulsive
case but in a reverse order with the top band corresponding to the
Mott states. By projecting the original Hamiltonian (1) into the
Hilbert spaces without any double occupancy, we can also get the
effective Hamiltonian given by Eq.(\ref{Heff}). Although the
effective Hamiltonian has the same form for both the repulsive and
attractive Hubbard model, the effective coupling strength $J$ has
different sign, i.e., $J=\pm \frac{4t^2}{|U|}$ for $U>0$ or $U<0$.
For $U>0$, the effective model is an AFM Heisenberg model which
describes the lowest continuum band of the original Hubbard model.
On the other hand, the effective model is a ferromagnetic (FM)
Heisenberg model for $U<0$, which describes the highest continuum
band of the attractive Hubbard model.

Given the initial state as the ground state of half-filling
repulsive Hubbard model (or effectively the ground state of AFM
Heisenberg model of Eq.(\ref{Heff}) labelled by the star in the
inset of Fig.7a), after a sudden switch of interaction from $U=15$
to $U=-15$, the state is transferred to the highest state of the
attractive Hubbard model (or effectively the highest excited state
of the FM Heisenberg model labelled by the star in the inset of
Fig.7b). By this way, we can effectively prepare the highest
excited state of a FM Heisenberg model. Here we indicate the
difference from the low-density case: the final state obtained by
sudden switch is on the top of the top band in Fig.7b, whereas the
FSTG state in Fig.3b is on the bottom of the top band. The
difference can be attributed to the different structures of the
energy spectra of the low-density and half-filling systems. In the
half-filling case, a narrow Mott band is formed with a Mott gap
separated from the excited bands.  The ground state and dynamic
behaviors are thus determined by the effective AFM Heisenberg
Hamiltonian, which describes the physics of the narrow Mott band
and displays different behaviors from the low-density case. On the
other hand, if the initial state is the ground state of
half-filling attractive Hubbard model, we can access the
repulsively paired state in the top band of the repulsive Hubbard
model by the sudden switch of the interaction to the repulsive
side.

\section{Summary}
%{\it Summary.}
In summary, we study the properties of the FSTG state of the
attractive Hubbard model, which is a highly excited state of the
attractive Hubbard model without paired states and corresponds to
the lowest real solution of the Bethe-ansatz equations with $U<0$.
Starting from the ground state of strongly repulsive spin-1/2
fermion in 1D deep optical lattices, such a state can be realized
via a sudden switch of the interaction to the strongly attractive
regime.  By calculating the transition probabilities, we have
shown that the excited FSTG state can be efficiently achieved in
the strongly interacting regime and thus is possible to be
realized experimentally with cold fermionic atoms in optical
lattices.

\begin{acknowledgments}
This work is supported by the NSF of China under Grant No.
10974234 and No. 11174360, National Program for Basic Research of
MOST and 973 grant.
\end{acknowledgments}

\end{document}